\documentclass[prl,twocolumn]{revtex4}   
\usepackage{graphicx}

\begin{document}

\title
{
Orbital-selective Mott transitions in the degenerate Hubbard model
}

\author
{                                                       
Akihisa Koga$^{1,2}$, Norio Kawakami$^1$, T. M. Rice$^2$ 
and Manfred Sigrist$^2$
}

\affiliation
{
$^1$Department of Applied Physics, Osaka University, 
Suita, Osaka 565-0871, Japan\\
$^2$Theoretische Physik, 
ETH-H\"onggerberg, Z\"urich 8093, Switzerland
}

\date{\today}

\begin{abstract}
 We investigate the Mott transitions in two-band Hubbard
models with different bandwidths. Applying dynamical mean field theory,
we discuss  the stability of itinerant quasi-particle states in each band.
We demonstrate that separate Mott transitions occur at different
Coulomb interaction strengths  
in general, which merge to a single transition only under special
conditions.  This kind of behavior 
may be relevant  
for the physics of the single-layer ruthenates, Ca$_{2-x}$Sr$_x$RuO$_4$.
\end{abstract}

\pacs{Valid PACS appear here}%

\maketitle

Strongly correlated multi-orbital electron
systems are among the most active topics in
condensed matter physics. In the Mott insulators the addition of
orbital to localized spin degrees of freedom leads to
complex ordered phase diagrams. In 
itinerant electron systems multiple Fermi surface sheets appear
with very distinct properties. Subtleties occur when
localized and itinerant electrons coexist, as is well known in the case of
itinerant $d$- and localized $f$-electrons which give rise to the rich
physics of heavy Fermion compounds. In view of their
very different character this is not so surprising. We may ask, however, 
whether it is possible to find coexistence of 
itinerant and localized
electrons for degenerate non-hybridizing
orbitals with small bandwidth difference.

A nearly degenerate $d$-electron system where multi-orbital properties
obviously play an important role is the single-layer isovalent 
ruthenate alloy
Ca$_{2-x}$Sr$_x$RuO$_4$. The end-member Sr$_2$RuO$_4$ is a well-known
unconventional superconductor \cite{PT,RMP}, while Ca$_2$RuO$_4$ is a
Mott-insulating $S=1$ antiferromagnet \cite{Nakatsuji1,Hotta}. The
relevant $4d$-orbitals belong to the $t_{2g}$-subshell.
The planar structure leads to very weak hybridization between orbital
which have even ($d_{xy}$) and odd parity ($d_{yz},d_{zx}$) under the
reflection $ z \to -z $. 
The complex evolution between these different end-members has 
led to various theoretical investigations \cite{Hotta,Fang}, and among
others to the proposal that some of
the $d$-orbitals display localized spin and orbital degrees of freedom,
and others provide itinerant electrons. This orbital-selective Mott
transition (OSMT) 
could explain the experimental
observation of a localized spin $S=1/2 $ in the metallic system at $x
\sim 0.5 $ which 
is difficult 
to obtain from the entirely itinerant ab initio description
\cite{Nakatsuji1,Anisimov,Fang}.  

The concept of the OSMT was recently challenged, 
in particular, by Liebsch whose dynamical mean field theory (DMFT) 
calculations suggested that two bands
of different width coupled by electron-electron interactions would
always undergo a common Mott transition \cite{Liebsch}. 
The aim of this paper is to revisit this problem and to analyze the
Mott transition in the degenerate two-orbital Hubbard model with
different bandwidths. 
Using another form of DMFT,  we find a different result and an
 OSMT. In addition, 
 we show also that correlations can stabilize a commensurate filling of one
 band even when the total electron count is fractional.

We consider the following Hubbard Hamiltonian with two orbitals,
\begin{eqnarray}
H&=&\sum_{\stackrel{<i,j>}{\alpha,\sigma}}
\left(t_{ij}^{(\alpha)}-\mu_{\alpha}\delta_{ij}\right)c_{i\alpha\sigma}^\dag
c_{j\alpha\sigma} 
+U\sum_{i\alpha}c_{i\alpha\uparrow}^\dag c_{i\alpha\uparrow}
c_{i\alpha\downarrow}^\dag c_{i\alpha\downarrow}\nonumber\\
&+&U'\sum_{i\sigma\sigma'}c_{i1\sigma}^\dag c_{i1\sigma}
c_{i2\sigma'}^\dag c_{i2\sigma'}
-J\sum_{i}c_{i1\sigma}^\dag c_{i1\sigma'}
c_{i2\sigma'}^\dag c_{i2\sigma}\nonumber\\
&-&J\sum_{i}\left[ c_{i1\uparrow}^\dag c_{i1\downarrow}^\dag
c_{i2\uparrow} c_{i2\downarrow}+H.c. \right]
\label{Hamilt}
\end{eqnarray}
where $c_{i\alpha\sigma}^\dag (c_{i\alpha\sigma})$ 
creates (annihilates) an electron 
with  spin $\sigma(=\uparrow, \downarrow)$ and orbital
index $\alpha(=1, 2)$ at the $i$th site. We restrict ourselves to the
case of non-hybridizing orbitals, relevant to the ruthenates, and
$t_{ij}^{(\alpha)}$ denotes the hopping
integral for orbital $ \alpha $,
$\mu$ the chemical potential,
$U$ ($U'$) the intraband (interband) Coulomb interaction
and $J$ the Hund coupling.
In the following, we restrict our discussions to the condition $U=U'+2J$,
obtained by symmetry arguments for degenerate orbitals.

We examine the stability of the metallic ground state of this model by means 
of DMFT which maps the lattice model to the problem of a single impurity
connected dynamically to a "heat bath" \cite{Georges}. 
The electron Green's function is obtained via the self-consistent
solution of this impurity problem. We represent the two electron bands by
semi-circular density of states (DOS),
$\rho_\alpha(x)=2/\pi D_\alpha \sqrt{1-(x/D_\alpha)^2}$
where $2D_\alpha$ is the bandwidth.
For the case of identical hopping integrals for the two bands,
$t_{ij}^{(\alpha)}=t_{ij}$ ($D_1=D_2$), 
 the role of orbital fluctuations has  been discussed 
by means of DMFT \cite{Degenerate,Koga}. There are various
methods to solve the effective impurity problem.
We use here the exact diagonalization method 
proposed by Caffarel and Krauth \cite{Caffarel}, 
since the quantum Monte Carlo (QMC) simulations used by Liebsch \cite{Liebsch} 
suffer from sign problems at low temperatures, in particular, if the Hund 
coupling is included. Additionally, we apply the linearized version of DMFT
(two-site DMFT) \cite{Potthoff},
which allows us to discuss electronic properties well even 
in the vicinity of the critical point. We restrict our discussions
to the paramagnetic case to clarify the nature of the Mott transition. 

\begin{figure}[htb]
\resizebox{0.32\textwidth}{!}{%
\includegraphics[width=6cm]{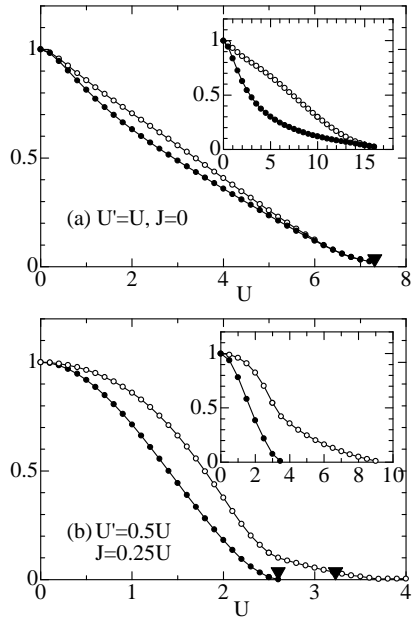}
}
\caption{The quasiparticle weights $Z_1$ and $Z_2$ at half filling
as a function of the Coulomb interaction $U$:
 (a) $U'/U=1.0$ ($J=0$)  and (b) $U'/U=0.5$ ($J/U=0.25$).  The
bandwidth is set as $D_1=1.0$ and $D_2=2.0$.
Open (closed) circles represent the results for orbital $\alpha=1(2)$
obtained by solving the DMFT impurity problem by means of the exact
diagonalization of a small cluster $(N=6)$.
Solid triangles represent the Mott-transition points
obtained by the two-site DMFT method, which produce the values
quite consistent with those of the numerical diagonalization.
Insets show the same plot for bandwidths
$D_1=1.0$ and $D_2=5.0$.
}
\label{fig:Z}       
\end{figure}

We first consider the case $ \mu_1 = \mu_2 =U/2+U'-J/2 $, i.e. both bands are
half filled.  
The quasiparticle weight $ Z_{\alpha}$, defined by
$Z^{-1}_\alpha=1-d \rm{Re}[\Sigma_\alpha(\omega)]/d\omega$
in terms of the self energy $\Sigma_\alpha(\omega)$ of each
band, will be used to characterize the stability of the metallic
state of the two bands.  
The results obtained with fixed ratios $U'/U$ and $J/U$
are shown in Fig. \ref{fig:Z} for half-filled bands.
We first focus on the case of $U=U'$ and $J=0$ [shown in
Fig. \ref{fig:Z} (a)]  with bandwidths $D_1=1.0$ and $D_2=2.0$.
When the Coulomb interaction is turned on, the 
quasiparticle weights $Z_1$ and $Z_2$ decrease
from unity in slightly
different ways reflecting the difference of the bandwidth.
A strong reduction of the quasiparticle weight 
appears  initially in the narrower band. 
However, when the system approaches the Mott transition, 
the quasiparticle weights merge again displaying a very similar
dependence on $ U $,  
and eventually reach zero at 
the same critical point.  The
inset shows the more extreme case of $D_1=1.0$ and $D_2=5.0$
(very wide band).
The common Mott transition originates from the enlarged symmetry
inherent in  $U=U'$ and $J=0$, which will be discussed below.
This result is in agreement with the conclusion
by Liebsch \cite{Liebsch}. For small interaction strengths the decrease
of the quasiparticle weight reflects the effective Coulomb interactions 
$U/D_\alpha$ which are different for two bands of different width,
$ D_{\alpha} $. The dependence of the initial decrease in $Z_{\alpha}$
on $U/D_\alpha$ reflects the general difference in $Z_{\alpha}$. 
On the other hand, in the vicinity of the quantum
critical point at the Mott transition the
effect of the bare bandwidth is diminished due to the
strong renormalization of the effective quasiparticle bandwidth
allowing $Z_1$ and $Z_2$ to vanish together
\cite{Gutzwiller,Brinkman}. 

The introduction of a finite Hund coupling $ J $ makes $ U \neq U' $
and leads to a qualitatively different behavior,
 as seen in Fig. \ref{fig:Z} (b).
With increasing $ U $ keeping the ratio $ U'/U=0.5 $ fixed, 
the quasi-particle weights decrease differently and vanish at different
critical points: $U_{c1}\approx 2.6$ for $ Z_1$ and $U_{c2} \approx
3.5 $ for $Z_2 $. Therefore, we observe an intermediate phase with one
orbital localized and the other itinerant, though strongly
renormalized ($ Z_2 \ll 1 $). The analogous behavior is observed for
different choices of the  
bandwidths, if $J$ takes a finite value [inset of Fig. \ref{fig:Z} (b)].
Although it is  difficult to precisely determine the second 
critical point $U_{c2}$,
this result certainly suggests the existence of 
the OSMT with $ U_{c2} > U_{c1} $.

\begin{figure}[htb]
\resizebox{0.32\textwidth}{!}{%
\includegraphics[width=6cm]{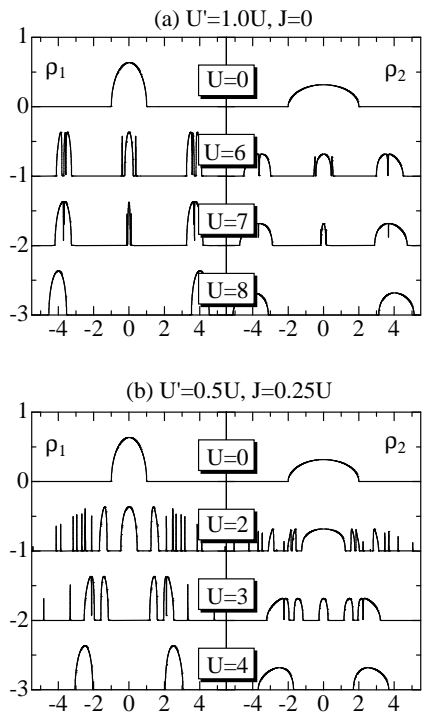}
}
\caption{The density of states $\rho_\alpha(\omega)$
at half filling: left (right) panel for $\alpha=1$ (2).
(a) $U'/U=1.0$ ($J=0$)  and (b) $U'/U=0.5$ ($J/U=0.25$).
It is clearly seen that the Mott transitions occur 
simultaneously in (a), while the orbital selective transitions 
occur in (b).
}
\label{fig:dos}       
\end{figure}

In Fig. \ref{fig:dos} we show how the quasiparticle states evolve
and then disappear  inside of the Mott-Hubbard gap. 
The DOS is computed by the two-site DMFT scheme \cite{Potthoff}.
In both cases   
the Mott-Hubbard gap develops as $U$ increases and
is accompanied by narrow quasiparticle mid-gap bands.
For case (a) with $J=0$, these quasiparticle bands
disappear simultaneously, whereas for case (b) with finite $J$,
they have different critical points,
consistent with the results mentioned above.

\begin{figure}[htb]
\resizebox{0.37\textwidth}{!}{%
\includegraphics{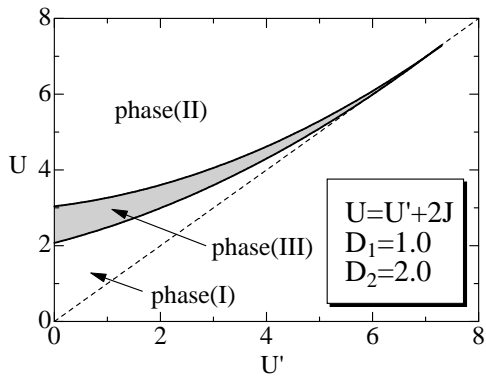}
}
\caption{Phase diagram for two-orbital Hubbard model
with  $D_1=1$ and $D_2=2$.  In the phase (I) [phase (II)], 
both bands are in  metallic (insulating) state.
The phase (III) is induced
by the orbital-selective Mott transition, where 
the metallic state coexists with the Mott insulating state.
Since we are concerned with the ferromagnetic Hund coupling, $J>0$,
the relevant region in the diagram is $U>U'$.
}
\label{fig:phase}       
\end{figure}

Repeating similar DMFT calculations for various 
choices of the parameters, we derive 
the ground-state phase diagram shown in Fig. \ref{fig:phase},
which displays some remarkable features.
First, the metallic phase (I) remains stable up to surprisingly 
large Coulomb interaction $U$ when $U \to U'$ (small $J$).
Here the Mott transitions merge to a single transition. 
This behavior originates from the high symmetry when
$ U=U' $ ($J=0$) with six degenerate two-electron onsite 
configurations: four spin configurations with one electron in each orbital
and two spin singlets with both electrons in one of the two orbitals.  
The additional symmetry in orbital/spin degrees
of freedom enlarges the phase space for charge fluctuations and leads to a
decrease of the  
Mott-Hubbard gap $ E_g $ at large $ U $. A rough estimate of $ E_g $
can be obtained 
from the second moment of the hopping Hamiltonian for a state $ | a
\rangle $ with an extra electron (or 
hole).  In $ \langle a | H^2 | a \rangle = T^2 $ all possible
configurations with the same onsite energies 
are considered as intermediate states \cite{Koch}. Because all charge
excitations mix with each other, there is only one gap. 
Assuming a staggered spin or orbital 
configuration as the most dominant local correlation 
for neighboring two-electron sites we obtain for the
effective hopping matrix element of an extra carrier
$ T =  \sqrt{t_1^2 + t_2^2} $. This
allows us to estimate the Mott-Hubbard gap at large $ U$:
\begin{equation}
E_{g} = U - 2z T =  U - 2z \sqrt{t_1^2 + t_2^2}
\end{equation}
where $ z$ is the coordination number. We can get a simple estimate of
$ U_c $ by setting 
$ E_g \to 0 $ leading to values $ U_c = 2D_1 \sqrt{2} $  (for  $ D_1 =
D_2 $) and 
$ U_c = 2 D_1 \sqrt{5} $ (for $ 2 D_1 = D_2 $ ).  Both estimates are
enhanced relative to the single-band case $ U_c = 2D $ \cite{Koga}. 

Away from the symmetric limit, i.e.  $ U > U' $ ($ 2J = U - U' $) orbital
fluctuations are suppressed and the spin 
sector is reduced by the Hund coupling to three onsite spin triplet
components as 
the lowest multiplet for two-electron sites. Applying the same
scheme as above, we recognize that charge excitations with two electrons 
in one or the other of the orbitals do not mix, since all hopping
processes included in (\ref{Hamilt}) preserve orbital 
configurations in the lowest multiplet sector. The effective hopping
for each orbital is now $ 
T_{\alpha} = t_{\alpha} / \sqrt{2} $ 
assuming a staggered spin-1 state at half-filling. 
The reduction compared to the single band case occurs due to the
locking of the spins into an onsite spin triplet. If we consider 
again the case $ 2 D_1 = D_2 $ we find two separate Mott transitions 
with critical values
\begin{equation}
U_{c1} =  \frac{1}{\sqrt{2}} D_1 + \frac{U'}{2}  \quad \mbox{and}
\quad U_{c2} = \sqrt{2} D_1 + \frac{U'}{2}
\;.
\end{equation}
In between the two transitions we find the metallic intermediate phase (III)
with one band localized and Mott-insulating and one band itinerant. 
Within our DMFT scheme we have also confirmed that various choices 
of bandwidths give rise to the qualitatively same structure of the
phase diagram  
as shown in Fig. \ref{fig:phase}.

Our result for the Mott transitions is different from that of both
Anisimov et al. and 
 Liebsch \cite{Anisimov,Liebsch}. The former group derived an
OSMT for
a special model which includes only the intraband Coulomb
repulsion $U$ within the DMFT approach, and drops effects
due to coupling between the orbitals. 
This scheme was criticized by Liebsch,
who took into account $U$, $U'$ and $J$. He claimed based on a DMFT
analysis that only a single Mott transition occurs in the generic
case. Liebsch's solution of the single-impurity problem within
DMFT is based on QMC and 
iterative perturbation methods. 
The former suffers from sign problems which limit its validity
at low temperatures, while the latter is an extrapolation from the small-$U$
regime. Our DMFT analysis which uses the exact diagonalization
of the impurity problem on a finite cluster is valid at zero-temperature and is
not restricted to weak coupling. Our results show separate Mott transitions
occurring generically except for the special case with high symmetry
$U=U'$ ($J=0$), for which the transitions merge irrespective of the
different bandwidths.

We have so far treated the case of two individually half-filled bands.
We now address the question what will happen when the electron count
is non-stoichiometric. 
This problem may provide another key  to understand the 
orbital selective Mott transitions 
in  $\rm Ca_{2-x}Sr_xRuO_4$ \cite{Nakatsuji1},
since each of the three original metallic bands possesses fractional fillings. 
In order to study this kind of system, we introduce a finite 
hole doping, and
observe how commensurability can emerge due to interactions.

\begin{figure}[htb]
\resizebox{0.32\textwidth}{!}{%
\includegraphics{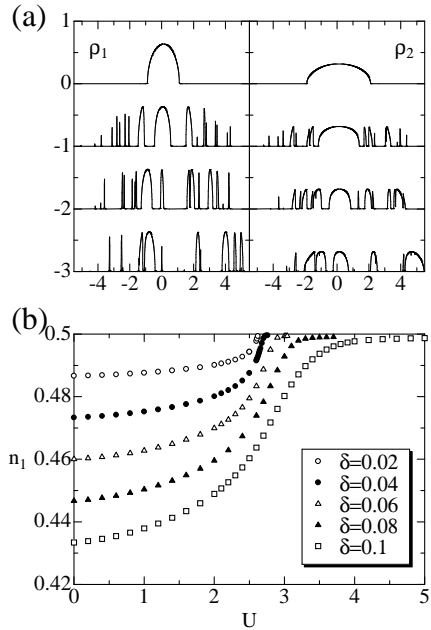}
}
\caption{
(a) The density of states $\rho_1(\omega)$ and
$\rho_2(\omega)$  for finite hole doping, $\delta=0.1$.
The Coulomb interaction is chosen 
as $U=0, 2.0, 3.0$ and $4.0$ 
$(U'=0.5U$ and $J=0.25U)$ from the top to the bottom.
(b) The number of electrons in the orbital $(\alpha=1)$ as a 
function of
$U$ 
when $U'=0.5U$ and $J=0.25U$.
}
\label{fig:doping}       
\end{figure}

In Fig. \ref{fig:doping} (a) we show the DOS
which is computed  by using  the two-site DMFT.
With increasing interactions quasiparticle states with large DOS
appear around the Fermi energy in both bands. Enhancing the   
interactions further we drive the first band insulating.
This is in contrast to
the single band system, where finite hole doping 
obscures the Mott transition and always gives metallic behavior. 
In the two-band system, however, commensurability in one of the bands
gradually emerges, as is clearly seen 
in  Fig. \ref{fig:doping} (b). The electron number for
the first band $n_1$ is plotted here.
When $U=0$, $n_1$ and $n_2(=1-\delta-n_1)$ are 
smaller than 0.5 because of finite hole-doping ($\delta=0.1$).
Coulomb interaction causes
electron transfer from one orbital to the other, giving rise to
one half-filled band at a certain interaction strength,
thereby causing an OSMT.

In conclusion we have discussed the Mott transitions 
in the degenerate Hubbard model with non-hybridizing orbitals which
have different bandwidths
by means of DMFT.
Our analysis has shown that a single Mott transition occurs 
when the Hund coupling is absent ($U = U'$), rendering the 
different bandwidth essentially irrelevant
at the transition point, as discussed by Liebsch.
In the more generic situation with finite Hund coupling,
 however, we find  the OSMT. 
This is also true for non-stoichiometric systems.
We believe that our study resolves the apparent
contradictions on this issue. Moreover it 
sheds light on the nature of such Mott transitions, which may be
relevant for the physics of the ruthenates.

\acknowledgments
We would like to thank V.I. Anisimov, Z. Fang, S. Nakatsuji, Y. Maeno and
M. Troyer for useful discussions.
This work was partly supported by a Grant-in-Aid from the Ministry 
of Education, Science, Sports and Culture of Japan 
and the Swiss National Science Foundation.
A part of computations was done at the Supercomputer Center at the 
Institute for Solid State Physics, University of Tokyo
and Yukawa Institute Computer Facility.

%


\end{document}